\documentclass[apj]{emulateapj}
\usepackage{graphicx}
\usepackage[]{natbib}
\usepackage{amsmath, multirow}
\usepackage{hyperref,breakurl}
\bibpunct{(}{)}{,}{a}{}{,}
\def\pasa{PASA}               
\begin{document}
\def\bullet{\object{1E0657$-$56}}
\def\bbullet{\object{MACS~J0025.4$-$1222}}
\def\macsos{MACS~J0717$+$3745}
\def\macsa{MACS~J0416.1$-$2403}
\def\macsb{MACS~J1149.5$+$2223}
\def\macsc{RXC~J2248.7$-$4431}
\def\HST{{\it HST}}
\def\Spitzer{\it Spitzer}
\def\arcsecf{\!\!^{\prime\prime}}
\def\arcminf{\!\!^{\prime}}
\def\diff{\mathrm{d}}
\def\ngx{N_{\mathrm{x}}}
\def\ngy{N_{\mathrm{y}}}
\def\eck#1{\left\lbrack #1 \right\rbrack}
\def\eckk#1{\bigl[ #1 \bigr]}
\def\round#1{\left( #1 \right)}
\def\abs#1{\left\vert #1 \right\vert}
\def\wave#1{\left\lbrace #1 \right\rbrace}
\def\ave#1{\left\langle #1 \right\rangle}
\def\kms{{\rm \:km\:s}^{-1}}
\def\dds{D_{\mathrm{ds}}}
\def\dd{D_{\mathrm{d}}}
\def\ds{D_{\mathrm{s}}}
\def\cs{\mbox{cm}^2\mbox{g}^{-1}}
\def\magz{m_{\rm z}}
\def\V{\rm{V_{\rm 606}}}
\def\ii{\rm{i_{\rm 775W}}} 
\def\iii{\rm{I_{\rm 814W}}} 
\def\z{\rm{z_{\rm 850LP}}}
\def\J{\rm{J_{\rm 110W}}}
\def\H{\rm{H_{\rm 160W}}}
\def\chone{3.6\:\mu\rm{m}}
\def\chtwo{4.5\:\mu\rm{m}}
\def\cunit{\mbox{erg}/\mbox{s}/\mbox{cm}^2/\mbox{\AA}}
\def\lunit{\mbox{erg}/\mbox{s}/\mbox{cm}^2}
\newcommand{\hii}{H{\sc ii}}
\newcommand{\Hii}{H~{\sc ii}}
\newcommand{\Ha}{\mbox{H$\alpha$}}
\newcommand{\hb}{\mbox{H$\beta$}}
\newcommand{\sii}{S~{\sc ii}}
\newcommand{\Feii}{Fe~{\sc ii}}
\newcommand{\oi}{O~{\sc i}}
\newcommand{\nii}{N~{\sc ii}}
\newcommand{\oiii}{O~{\sc iii}}
\newcommand{\mgii}{Mg~{\sc ii}}
\newcommand{\tco}{{$^{13}$CO}}
\newcommand{\CO}{{$^{12}$CO}}
\newcommand{\Tco}{{$^{12}$CO}}
\newcommand{\co}{C{$^{18}$O}}
\def\ct{[\ion{C}{2}]}
\def\cte{\rm [C \sc{II}]}
\def\ot{[\ion{O}{3}] 88$~\mu\mbox{m}$}
\def\ctf{[\ion{C}{2}] 158$~\mu\mbox{m}$}
\def\lya{Lyman-$\alpha$}
\def\ha{H$\alpha$}
\newcommand{\hst}{\it HST}
\newcommand{\galfit}{\texttt{GALFIT}}    
\newcommand{\mopex}{\texttt{mopex}}      
\newcommand{\surfsup}{SURFS UP}         
\newcommand{\sex}{\texttt{SExtractor}}   
\newcommand{\sersic}{S\'ersic}           
\newcommand{\pygfit}{\texttt{PyGFIT}}    
\newcommand{\lephare}{\texttt{Le Phare}} 
\newcommand{\eqn}[1]{equation~(\ref{#1})}
\newcommand{\fig}[1]{Figure~\ref{#1}}
\newcommand{\tab}[1]{Table~\ref{#1}}


\title{ALMA {\ctf} detection of a redshift 7 lensed galaxy behind RXJ1347.1$-$1145 \altaffilmark{*}}
\altaffiltext{*}{These observations are based on the following ALMA data:
ADS/JAO.ALMA\#2015.1.00091.S. They are also associated with
programs {\Spitzer} \#90009, 60034, 00083, 50610, 03550, 40593, and
{\hst} \# GO10492, GO11591, GO12104, and GO13459. Furthermore based on multi-year KECK programs.}
\shorttitle{}
\author{Maru\v{s}a Brada\v{c}\altaffilmark{1},
Diego Garcia-Appadoo \altaffilmark{2,3},
Kuang-Han Huang\altaffilmark{1},
Livia Vallini\altaffilmark{4,5,6},
Emily Quinn Finney\altaffilmark{1},
Austin Hoag\altaffilmark{1},
Brian \ C. Lemaux\altaffilmark{1},
Kasper Borello Schmidt\altaffilmark{7},
Tommaso Treu\altaffilmark{8,x},
Chris Carilli\altaffilmark{9,10},
Mark Dijkstra\altaffilmark{11},
Andrea Ferrara\altaffilmark{12,13},
Adriano Fontana\altaffilmark{14},
Tucker Jones\altaffilmark{1},
Russell Ryan\altaffilmark{15},
Jeff Wagg\altaffilmark{16},
Anthony \ H. Gonzalez\altaffilmark{17}
}
\shortauthors{Brada\v{c} et al.}
\altaffiltext{1}{Department of Physics, University of California, Davis, CA 95616, USA}
\altaffiltext{2}{Joint ALMA Observatory, Alonso de C\'{o}rdova 3107, Vitacura, Santiago, Chile}
\altaffiltext{4}{European Southern Observatory, Alonso de C\'{o}rdova
  3107, Vitacura, Santiago, Chile}
\altaffiltext{4}{Nordita, KTH Royal Institute of Technology and Stockholm University, Roslagstullsbacken 23, SE-106 91 Stockholm, Sweden}
\altaffiltext{5}{Dipartimento di Fisica e Astronomia, viale Berti Pichat 6, I-40127 Bologna, Italy}
\altaffiltext{6}{INAF, Osservatorio Astronomico di Bologna, via Ranzani 1, I-40127 Bologna, Italy}
\altaffiltext{7}{Leibniz-Institut f\"{u}r Astrophysik Potsdam (AIP), An der Sternwarte 16, 14482 Potsdam, Germany}
\altaffiltext{8}{Department of Physics and Astronomy, UCLA, Los Angeles, CA, 90095-1547, USA}
\altaffiltext{9}{National Radio Astronomy Observatory, P. O. Box 0, Socorro, NM 87801, USA}
\altaffiltext{10}{Cavendish Laboratory, 19 J. J. Thomson Ave., Cambridge CB3 0HE, UK}

\altaffiltext{11}{Institute of Theoretical Astrophysics, University of Oslo, Postboks 1029 Blindern, NO-0315 Oslo, Norway}
\altaffiltext{12}{Scuola Normale Superiore, Piazza dei Cavalieri 7, I-56126 Pisa, Italy}
\altaffiltext{13}{Kavli IPMU (WPI), Todai Institutes for Advanced Study, the University of Tokyo, Japan}
\altaffiltext{14}{INAF - Osservatorio Astronomico di Roma Via Frascati 33 - 00040 Monte Porzio Catone, 00040 Rome, Italy}
\altaffiltext{15}{Space Telescope Science Institute, 3700 San Martin
  Drive, Baltimore, MD 21218, USA}
\altaffiltext{16}{Square Kilometre Array Organisation, Lower
  Withington, Cheshire, UK}
\altaffiltext{17}{Department of Astronomy, University of Florida, 211 Bryant Space Science Center, Gainesville, FL 32611, USA}
\altaffiltext{x}{Packard Fellow}
\email{marusa@physics.ucdavis.edu}


\begin{abstract}  
  We present the results of ALMA spectroscopic follow-up of a
  $z=6.766$ {\lya} emitting galaxy behind the cluster RXJ1347.1$-$1145. We
  report the detection of {\ctf} line fully consistent with the {\lya}
  redshift and with the peak of the optical emission. Given the
  magnification of $\mu=5.0 \pm 0.3$, the intrinsic (corrected for
  lensing) luminosity of the {\mbox{\ct}} line is $L_{\cte}
  =1.4^{+0.2}_{-0.3} \times 10^7L_{\odot}$, roughly ${\sim}5$ times
  fainter than other detections of $z\sim 7$ galaxies. The result
  indicates that low $L_{\cte}$ in $z\sim 7$ galaxies compared to the
  local counterparts might be caused by their low metallicities and/or feedback. The small velocity off-set ($\Delta v = 20_{-40}^{+140}\kms$) between the {\lya} and {\ct} line is unusual, and may be indicative of ionizing photons escaping.  
\end{abstract}

\section{Introduction}
\label{sec:intro}

The epoch of reionization, during which the Universe became
transparent to UV radiation, is still poorly understood. Faint galaxies are likely responsible for this
transformation, however this connection is far from confirmed (e.g. \citealp{robertson15,madau15}). Despite
great progress in finding candidates with  Hubble Space
Telescope {\hst}
(e.g. \citealp{bouwens15}) beyond $z\gtrsim 7$, galaxies remain
enshrouded in mystery, at least from a spectroscopic point of
view. Spectroscopic confirmations remain extremely difficult%
, as the most prominent spectral feature
in the optical/near IR wavelengths, the {\lya} line, can be erased
by neutral gas both in and surrounding galaxies.

An alternative way of measuring redshifts for early galaxies is to use sensitive
radio/far-infra-red (FIR) telescopes to observe the {\ct}
$158\mu\mbox{m}$ line. It is among the strongest lines in star-forming
galaxies at radio through FIR wavelengths (e.g., \citealp{carilli13}) and it is
therefore being actively pursued as a new way to measure redshifts at
$z\gtrsim 6$.  Furthermore, for lower redshift galaxies there exists a
  connection of {\ct} to star formation rate
  (SFR, e.g., \citealp{delooze14}), although for $z\sim 7$ such a relation
  has not been studied in detail yet due to limited sample
  size. {\ct} is also a very useful tracer of  the kinematics of distant galaxies. 
However, even with ALMA, detections at $z\sim 7$
remain sparse (see \citealp{pentericci16,maiolino15,knudsen16a} for
current detections of {\ct} and \citealp{inoue16} for an {\ot} detection). In particular, despite deep observations, the extremely
bright {\lya} Emitter (LAE) Himiko was not detected in {\ct}
\citep{ouchi13,ota14,schaerer15}.

From the theoretical side, the issue of non-detections has been
studied e.g., by \citet{vallini15,olsen15,narayanan16}. Both
\citet{vallini15} and \citet{olsen15} conclude that in high-z galaxies the {\ct}
emission arises predominately from photodissociation regions
(PDRs). Furthermore, at $z\sim 7$ the deficit in {\ct}-emission at a given SFR has been ascribed to either negative stellar
feedback disrupting molecular clouds near the star forming regions
and/or low gas metallicities. \citet{narayanan16} point out that also
the cloud surface density is a key parameter, with the {\ct}
luminosity decreasing with decreasing size of  the {\ct}
emitting region in high surface density clouds. Other factors that
influence total {\ct} emission include (i) the relative
abundances of the various gas phases composing the interstellar medium
ISM (ionized,
neutral and molecular), (ii) the hardness of the radiation field,
and (iii) the temperature and density of the emitting gas.

In this paper we report on ALMA observations of  RXJ1347:1216, a
normal ($L < L^*$) star forming galaxy with {\lya} emission at
$z_{\rm Ly\alpha}=6.7659 ^{+0.0030}_{-0.0005}$. It was first reported by
\citet{bradley14} and \citet{smit14} as a
photometrically selected $z\sim 7$ galaxy from the {\hst} and {\Spitzer} CLASH data (Cluster Lensing And
Supernova survey with Hubble, \citealp{postman12}). We have
spectroscopically confirmed its redshift using Keck DEIMOS data
\citep{huang15a} and {\hst} grism data from GLASS
\citep{treu15,schmidt16}. In \citet{huang15a} we also  measured its
integrated stellar properties
using deep {\Spitzer} SURFSUP data (Spitzer UltRa Faint SUrvey Program, \citealp{surfsup}). Our magnification model shows that the massive foreground galaxy cluster RX J1347.1$-$1145 magnifies
 RXJ1347:1216 by a factor $5.0\pm 0.3$. Taking into account
 magnification, the galaxy's intrinsic rest-frame UV luminosity is
 $0.18^{+0.07}_{-0.05} L^*$ (assuming characteristic magnitude at $z\sim 7$ of $M_{UV}^*=-20.87\pm 0.26$ from
\citealp{bouwens15})  making
RXJ1347:1216 the first galaxy detected with ALMA at $z\sim 7$ having
a luminosity characteristic of the majority of galaxies
at $z\sim 7$.


Throughout the paper we assume a
$\Lambda$CDM concordance cosmology with $\Omega_{\rm m}=0.27$,
$\Omega_{\Lambda}=0.73$, and Hubble constant
$H_0=73{\rm\ kms^{-1}\:\mbox{Mpc}^{-1}}$.
Coordinates are given for the epoch J2000.0, and magnitudes are in the
AB system.

\section{Observations and data reduction} \label{sec:datared}
We observed RXJ1347:1216 with ALMA on July 21 2016 in Band 6 with
  38 12-m antennae on a configuration of 15-700m baselines. The precipitable
  water vapor stayed stable at $\sim$ 0.8 mm during the
  observations. The total time on-source was 74 minutes, with the
  phases centered at the {\hst} position of the source. Out of the
  four spectral windows, SPW0 was set to Frequency Division Mode and
  its center tuned to the {\ctf} rest-frame frequency of $1900.54
  \mbox{GHz}$ and a sky-frequency of $244.85 \mbox{GHz}$, in the Upper
  Side Band, yielding a velocity resolution of 9.5 $\kms$ after a spectral averaging factor of 16 was applied to reduce the data rate.  The other three spectral windows were used for continuum in Time Division Mode ($31.25\mbox{MHz}$  spectral resolution) at lower frequencies. We used J1337$-$1257 for bandpass and absolute flux scale calibrators and J1354$-$1041 for a phase calibrator. 

The data reduction followed the standard procedures in the Common
Astronomy Software Applications (CASA) package. The data cube was
cleaned using Briggs weighting and ROBUST $=$ 0.5. The FWHM beam size
of the final image is $0.58\arcsec \times 0.41\arcsec$ at a position
angle of $288\deg$.  The $1\sigma$ noise of the {\ctf} line image is
$\sigma_{\rm line} = 250\mu \mbox{Jy beam}^{-1}$ at $244.7424
\mbox{GHz}$ over a channel width of 30$\kms$ ($24.5\mbox{MHz}$). The
continuum image was extracted using all the line-free channels of the
four spectral windows, resulting in a continuum sensitivity of $<15\mu
\mbox{Jy}$ ($1\sigma$). Flux calibration errors ($\sim 5\%$) are included in all measurements.

\begin{deluxetable}{lc}  
 \tablecolumns{2}
\tablewidth{0pc}
\tablecaption{Stellar population modeling results for RXJ 1347:1216 using HST and Spitzer photometry from
  \citet{huang15a}, spectroscopy of {\lya} from Keck DEMOS from
  \citet{huang15a} and {\ct} from ALMA observations. \label{tab:prop}}
\startdata 
\hline
\multicolumn{2}{c}{ALMA}\\
R.A.  &  13:47:36.214\\
Dec. & $-$11:45:15.20\\
$z_{\cte}$ &$6.7655\pm 0.0005$\\
$S_{\rm line}$ & $1.25\pm 0.25 \mbox{mJy}$ \\
$S_{\rm line,g}$ (Gauss Fit) & $0.82\pm 0.26 \mbox{mJy}$ \\
$S_{\rm line}\Delta v $ (Gauss Fit) & $67\pm 12\mbox{mJy kms}^{-1}$\\
FWHM (Gauss Fit) & $75 \pm 25 \mbox{kms}^{-1}$\\
$L_{\cte} \times f_{\mu}\tablenotemark{b} $& $1.5^{+0.2}_{-0.4} \times
10^7L_{\odot}$\\
Continuum &  $<15\mu \mbox{Jy}$ ($1\sigma)$\\
$L_{\rm FIR}\times f_{\mu}\tablenotemark{b}$ & $<2\times 10^{10}L_\odot$ ($3\sigma$)\\
$SFR_{\rm FIR}\times f_{\mu}\tablenotemark{b}$ & $<3M_\odot\,\text{yr}^{-1}$ ($3\sigma$)\\
\hline
\multicolumn{2}{c}{HST + Spitzer + Keck}\\
R.A.  &  13:47:36.207\\
Dec.& $-$11:45:15.16\\
$z_{\rm Ly\alpha}$ &$6.7659 ^{+0.0030}_{-0.0005}$\\
$EW_{\rm Ly\alpha}$ & $26\pm 4\mbox{\AA}$\\
$\Delta_{\rm HST-ALMA}[\mbox{arcsec} (\mbox{kpc})]$ & 0.1 (0.5)\\
 $\Delta v_{Ly\alpha-\cte} $ &$  20_{-40}^{+140}\kms$\\
F160W\tablenotemark{a} & $26.1\pm 0.2 \mbox{mag}$\\
 $\mu_0$ &$5.0 \pm 0.3$\\
 $M_{\text{stellar}} \times f_{\mu}\tablenotemark{b}$  &$8.0^{+6.5}_{-0.9} \times 10^7~M_\odot$ \\
 $SFR_{\rm SED} \times f_{\mu}\tablenotemark{b}$& $8.5^{+5.8}_{-1.0}~M_\odot\,\text{yr}^{-1}$\\
 $SFR_{\rm UV} \times f_{\mu}\tablenotemark{b}$& $3.2\pm 0.4~M_\odot\,\text{yr}^{-1}$\\
Age & $\leq 13$Myr \\
$sSFR $ & $105.1^{+0.1}_{-20} \mbox{Gyr}^{-1}$\\
$E(B-V)_{\text{fit}}$ &$0.10^{+0.05}_{-0.01}$\\
$\beta_{\rm UV}$ &$-2.5^{+0.7}_{-1.0}$\\
\enddata
\tablenotetext{a}{Lensed total magnitude ({\tt MAG\_AUTO} as defined
  by SExtractor \citealp{sextractor}) in F160W.}
\tablenotetext{b}{The intrinsic properties calculated assuming $\mu = \mu_0$
  from this table. To use a different magnification factor $\mu$,
  simply use $f_{\mu} \equiv \mu/\mu_0$, i.e. dividing value given in this row by $f_{\mu}$.}
\end{deluxetable}

\section{Results} \label{sec:results} 
\begin{figure}[h!]
\centerline{\includegraphics[width=0.5\textwidth]{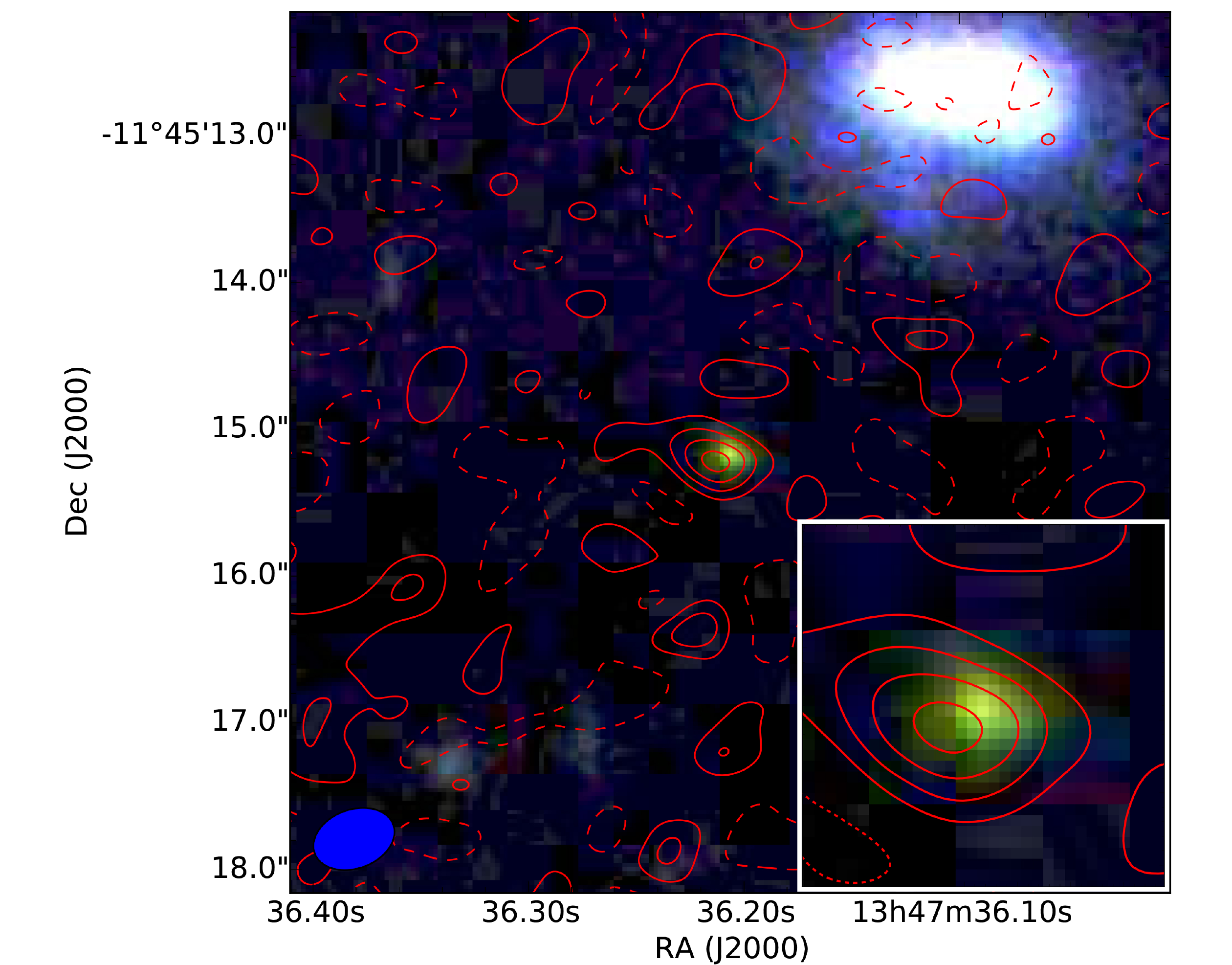}}
\caption{{\ct} emission overlaid on the HST color composite RGB image
  (optical-F110W-F160W). The contours are spaced linearly between $1-5\sigma$ (solid lines); negative contours ($-1,-2\sigma$) are given as dashed lines. A $1\arcsec\times1\arcsec$ zoom-in is shown in the inset, and the beam is given in bottom-left. \label{fig:cii}}
 \end{figure}

\begin{figure}[h!]
\centerline{\includegraphics[width=0.5\textwidth]{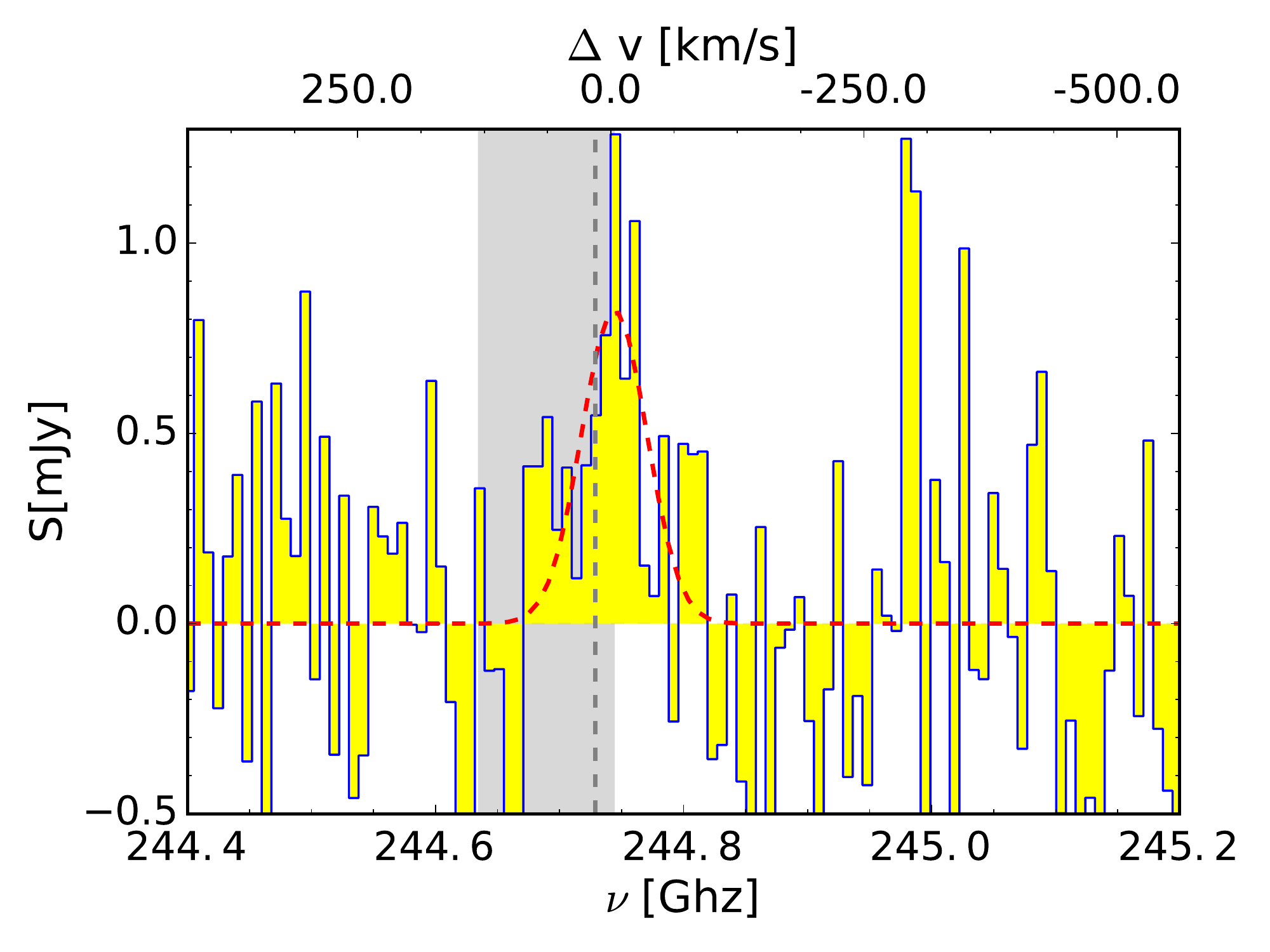}}
\caption{Extracted spectrum (flux $S$ as a function of frequency
  $\nu$) of the {\ct} emission. The red line
  denotes the best fit Gaussian (with parameter listed in
  Table~\ref{tab:prop}) and the grey dashed line and region correspond
  to the {\lya} redshift and uncertainty. \label{fig:spectrum}}
 \end{figure}

We have detected {\ct} emission
in RXJ1347:1216 with {\lya} emission first reported by
\citet{huang15a} and \citet{schmidt16}. The {\ct} line is detected at
$5\sigma$ (peak line flux $S_{\rm line} = 1.25\pm 0.25 \mbox{mJy}$
using $30\kms$ resolution, Fig.~\ref{fig:cii}).  Due to gravitational
lensing, we are able to measure {\ct} luminosity in RXJ1347:1216 that
is intrinsically ${\sim}5$ times fainter than other such measurements
at $z\sim 7$ to date (and similar luminosity to a $z\sim 6$ object in
\citealp{knudsen16}). We extract the spectrum using native spectral
resolution with channel width of $9.6~\kms$ ($7.8125\mbox{MHz}$) and
measure {\ct} redshift of $z_{\cte} = 6.7655\pm 0.0005$
(Fig.~\ref{fig:spectrum}). We fit the line using a Gaussian and
estimate peak flux of $S_{\rm line,g} = 0.82\pm 0.26 \mbox{mJy}$, FWHM
of $75 \pm 25~\kms$ and the integrated line flux of $S_{\rm
  line}\Delta v = 67\pm 12\mbox{mJy kms}^{-1}$. The integrated values do not critically depend
upon the assumption of a Gaussian profile within the uncertainties. The luminosity is
$L_{\cte} = 1.5^{+0.2}_{-0.4} \times 10^7L_{\odot}$, corrected for
lensing and the errors reflect both uncertainties in the flux estimates as well as magnification
(Table~\ref{tab:prop}).

\begin{figure}[h!]
\centerline{\includegraphics[width=0.55\textwidth]{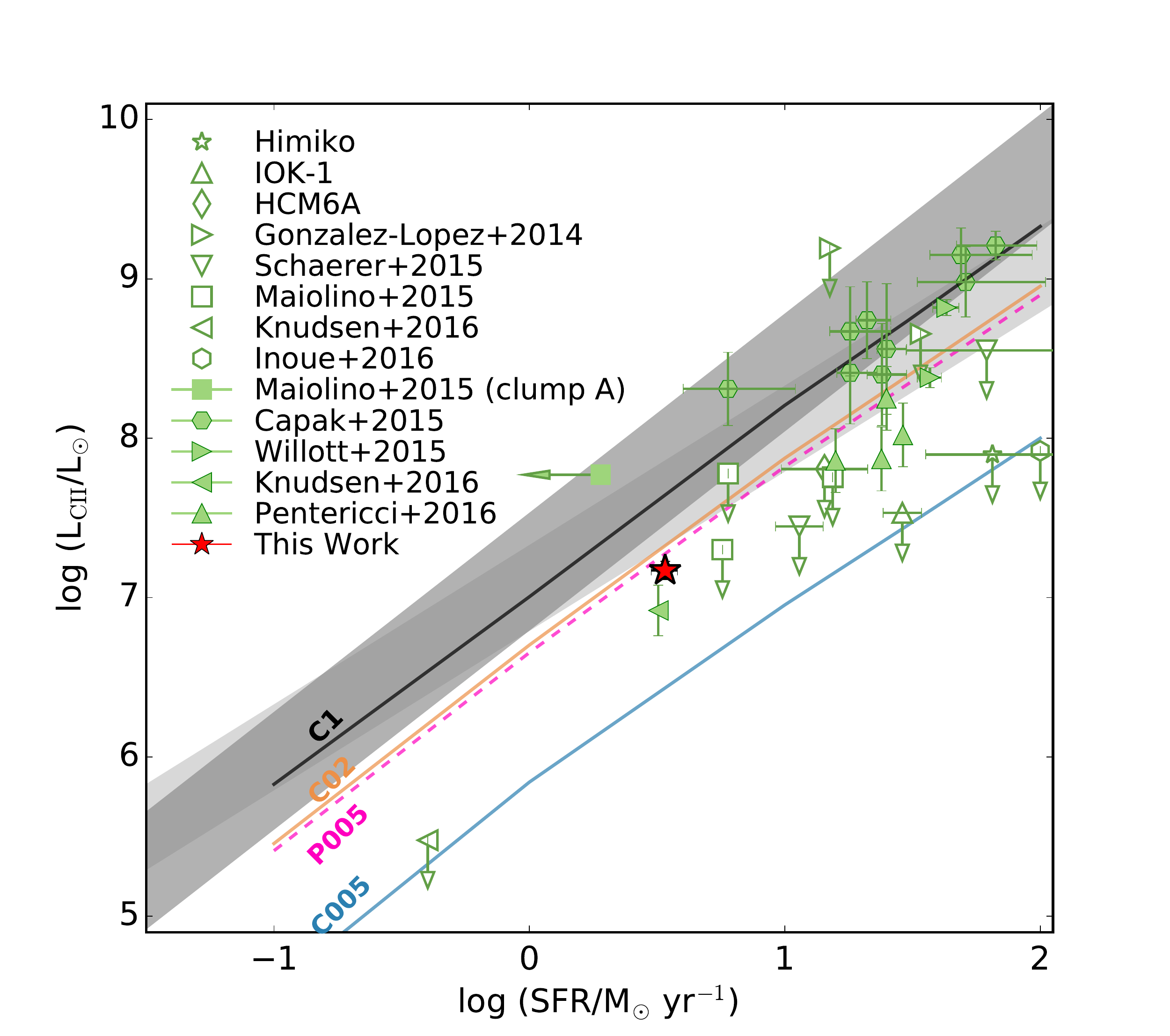}}
\caption{$L_{\cte}$ vs SFR at $z>5$.  The red
star is the {\ct} detection of the object RXJ 1347:1216, we show the SFR estimated directly from
 UV luminosity $\rm{SFR}_{\rm UV}$.
Lines represent the result from \citet{vallini15} obtained assuming a
constant metallicity: black for $Z = Z_{\odot}$ (C1), orange for $Z = 0.2Z_{\odot}$ (C02), and
blue for  $Z = 0.05Z_{\odot}$ (C005). The magenta dashed line (P005) corresponds to a
density-dependent metallicity with  $Z = 0.05Z_{\odot}$. The 1$\sigma$ scatter around the best-fit relations
for dwarf and local starburst galaxies from \citet{delooze14} are plotted in dark gray and light gray,
respectively. The green empty (filled) points represent upper limits (detections)
of {\ct} in {\lya} and Lymann break galaxies at redshifts $5-7$ as reported by
\citet{pentericci16,knudsen16,inoue16,maiolino15,capak15,willot15,schaerer15,gonzalezlopez14,ota14,ouchi13,kanekar13}. We excluded detections in more extreme objects (quasars and starbursts) for the clarity of the plot. \label{fig:ciisfr}}
 \end{figure}

We investigate the $L_{\cte}$-SFR connection for this galaxy. The
strength of the {\ct} emission correlates with the SFR in local dwarf galaxies \citep{delooze14},
 however the {\ct} luminosity is not a simple function of SFR, nor is
 there a simple correlation between {\ct} luminosity and the total mass
 of the ISM at higher redshifts. This is shown in Fig.~\ref{fig:ciisfr} where we plot $L_{\cte}$ vs. SFR. We include results from simulations by
 \citet{vallini15} as well as results for $z>5$ galaxies from the
 literature. We show the SFR of RXJ1347:1216 estimated directly from
 UV luminosity $\rm{SFR}_{\rm UV} = 3.2\pm
 0.4~M_\odot\,\text{yr}^{-1}$  in Fig.~\ref{fig:ciisfr}, assuming no
 dust attenuation, to compare with values from the literature.

Using ALMA FIR continuum non-detection limits ($<15\mu \mbox{Jy}$) we derive
$\rm{SFR}_{\rm{FIR}}<3~M_{\odot}~\rm{yr}^{-1}$ (assuming spectral index $\beta = 1.5$ and conservatively a dust
 temperature of $45~\mbox{K}$). This is in contrast to our nominal
 SED-fitting results from \citet{huang15a} using $0.2\,Z_\odot$
 templates from \citet{bc03} and assuming \citet{calzetti00} dust
 attenuation curve which imply a dust-obscured
 $\rm{SFR}_{\rm{SED}}\approx14~M_{\odot}~\rm{yr}^{-1}$, significantly higher than
 the FIR upper limit. Re-fitting the SED assuming a constant star
 formation history and a steeper SMC dust attenuation curve
 \citep{pei92}, motivated by recent ALMA results (e.g., \citealp{capak15}),
 leads to a smaller dust-obscured
 $\rm{SFR}_{\rm{SED}}=5.3~M_{\odot}~\rm{yr}^{-1}$, closer to but still
 higher than the limit derived from the non-detection of FIR
 continuum. This is similar to $z\sim 5$ sources in \citet{capak15}. 

\begin{figure}[h!]
\centerline{\includegraphics[width=0.5\textwidth]{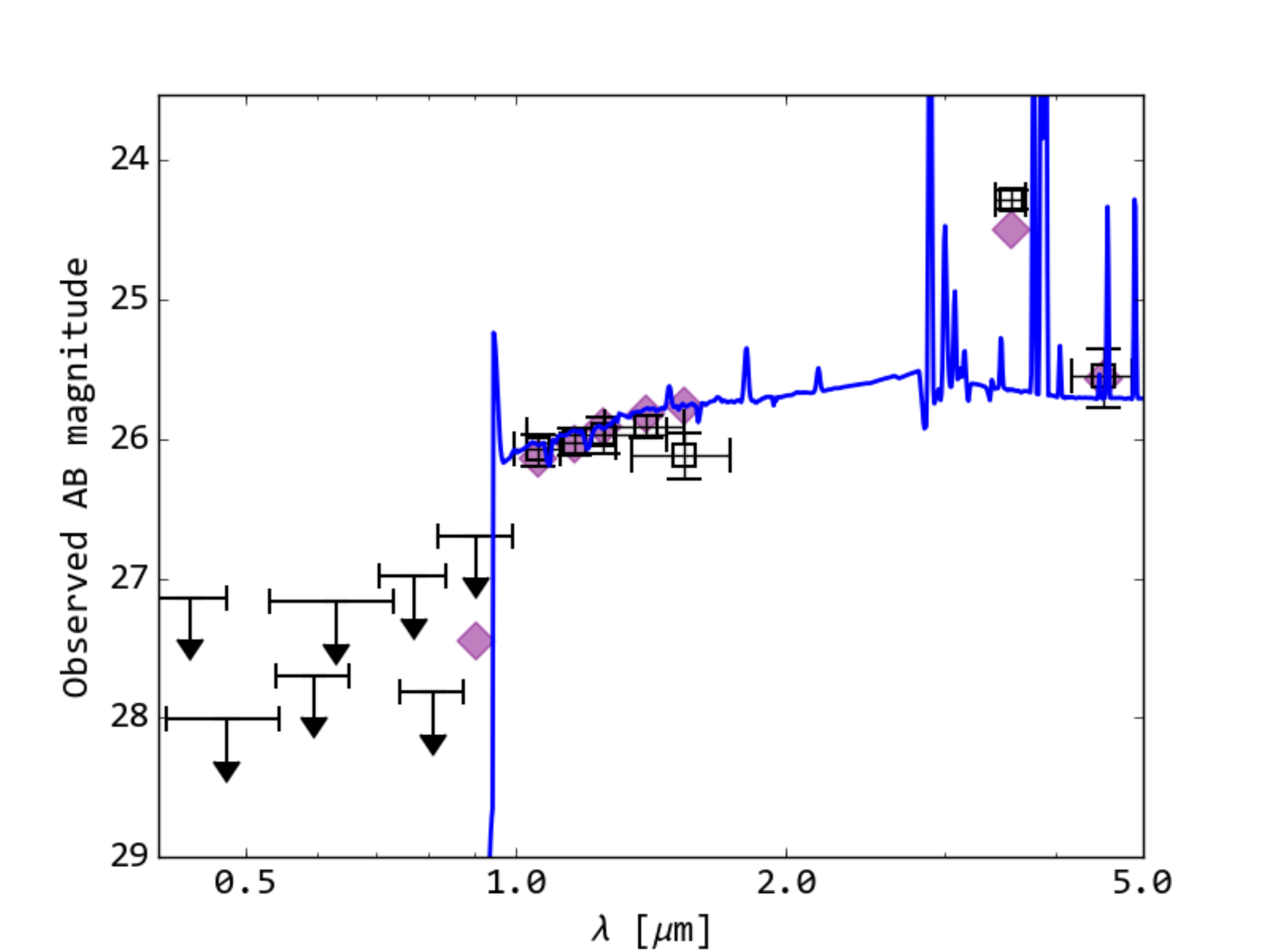}}
\caption{SED fitting of HST+Spitzer photometry \citep{huang15a}, using SMC dust attenuation curve from \citet{pei92}. We assume a metallicity
  that is $Z=0.2Z_{\odot}$, constant star formation history, and we
  fix the templates at $z=6.7655$. Squares/upper limits represent
  measured photometry from \citet{huang15a}, while purple diamonds
  indicate the model (blue line) predicted magnitudes in each band. \label{fig:sed}}
 \end{figure}

The nominal UV slope $\beta_{\rm UV}$ measured from the observed (i.e. not fitted) 
 $\mbox{F125W}$ to ${\mbox F160W}$ magnitudes (see \citealp{huang15a} for
 details) is quite blue ($\beta_{\rm
   UV}= -2.5^{+0.7}_{-1.0}$) with large uncertainty. 
 The stellar mass is $M_{\text{stellar}} =
 8.0^{+6.5}_{-0.9} \times 10^7~M_\odot$ and young age of  $\leq 13$Myr  (with a
 maximally-old stellar population not contributing $>10\%$ of the
 stellar mass).  We list the SED fitting results using the
 SMC curve in Table~\ref{tab:prop} and show the best-fit template in Figure~\ref{fig:sed}. Most of the
 changes  relative to \citet{huang15a} come from the difference in dust attenuation curve; changing
 the star formation history alone changes the results by less than
 10\%.

 The low {\ct} luminosity given SFR is
 consistent with models requiring low metallicity $\lesssim 0.2Z_{\sun}$,
 though feedback might play a role as well.  As noted by
 \citet{huang15a} and \citet{smit14}, SED modelling for this galaxy
 (Fig.~\ref{fig:sed}) requires strong nebular emission lines to
 explain its observed IRAC $[3.6]-[4.5]$ color. The object is best
 fit with a young ($\leq 13\mbox{Myr}$) stellar population, low metallicity
 ($0.2Z_{\odot}$), and strong nebular emission lines
 ($EW>1000\mbox{\AA}$ for {\hb} and [OIII], though note that all these
 parameters are highly degenerate), giving a consistent picture with
 lower {\ct} luminosity compared to the local galaxies.  The recent detection of {\ot} in a
 galaxy at $z = 7.2120$ with ALMA by \citet{inoue16} along with the
 upper limit on the {\ct} is also  in line with our results as it requires very low gas-phase metallicities (\citealp{vallini16}, see also \citealt{cormier15} for a study of local dwarf
 galaxies with high {\ot}/{\ct} ratios).

 Following \citet{pentericci16} and \citet{wang13}, we can also estimate the
 dynamical mass within the {\ct}-emitting region of the galaxy. We use $M_{\rm dyn} =
 1.16\times10^5V^2_{\rm cir}D$, 
where $V_{\rm{cir}}$ is the circular velocity in $\kms$, estimated
using $V_{\rm{cir}} = 0.75\times\rm{FWHM}_{\rm{[CII]}}/\sin(i)$, $i$
is the disk inclination angle, and $D$ is the disk diameter in kpc. Note that this approximation is very uncertain, as $z\sim 7$
 galaxies might not have ordered motion. However, \citet{pallottini16}
 showed that {\ct} mostly arises from an ordered,
 albeit small, disk. We estimate the disk diameter to be $\sim 0\farcs
 6$ and
 obtain $M_{\rm dyn} \sin ^2 i = (1.0\pm 0.5) \times 10^9
 M_{\sun}$. For an intrinsically axisymmetric, infinitesimally thin
 disk, inclination is related to ellipticity via $\cos i = b/a$, where
 $b/a$ is the axis ratio. 
With $b/a=0.6$ measured from the {\ct} emission, we get $i=50^{\circ}$ and $M_{\rm{dyn}}=(2.0\pm0.9)\times10^9\,M_\odot$ (the error does not include the systematic errors from inclination). This value indicates that $f_{\rm gas}\sim 1-M^{*}/M_{\rm dyn}=96^{+2}_{-10}\%$
 of the baryonic mass of this galaxy is in the molecular and atomic gas (contribution of dark matter to $M_{\rm{dyn}}$  within the {\ct}-emitting region is small). The
 high value is consistent with predictions from the Kennicutt-Schmidt
 relation of SFR to gas mass \citep{kennicutt98}, from which we obtain
 $M_{\rm gas}=(3.5^{+2.1}_{-1.1}) \times 10^9 M_{\sun}$ (using value $\rm SFR_{\rm SED}$ from Table~\ref{tab:prop}).

The {\ct} line is spatially consistent (measured offset is $0\farcs 1$) with the
rest-frame UV emission within the standard ALMA astrometric
uncertainty of $\sim
0\farcs1$ ({\hst} uncertainty is smaller). In addition, \citet{dunlop16} noted that the HST and ALMA astrometry
of the HUDF presented both a systematic shift of $0\farcs 25$
and a random offset of up to $0\farcs5$. Similarly,
\citet{pentericci16} notes $0\farcs 1-0\farcs 6$ random offsets measured
from serendipitous detections in the field. Unfortunately, we do not
detect any other sources in our small field-of-view to perform
relative astrometric calibration. 

 The {\lya} line redshift of this object is also in excellent agreement with
 the {\ct} redshift (Fig.~\ref{fig:spectrum}).  The resulting velocity
 offset of {\lya} compared to {\ct} is only $\Delta v =
 20_{-40}^{+140}\kms$ (68\% confidence, positive $\Delta v$ indicates that
 {\lya} is redshifted). The  {\lya} redshift was difficult to determine
 with a high accuracy given the proximity of a skyline, hence in
 \citet{huang15a} we reported it only with two significant digits. We
 remeasure the redshift using our DEIMOS data (and improve on absolute
 wavelength calibration reported in \citealp{huang15a}). The reasons
 for asymmetric errors on the measurement (Table~\ref{tab:prop}) are
 the proximity of the skyline, lower S/N of the line ($\sim10\sigma$), and the asymmetric nature of the line. The errors are, however, small and we do not
 detect significant shift of the {\lya} line.

 This is somewhat unexpected, as for such a low
 luminosity galaxy ($0.18^{+0.07}_{-0.05} L^*$, rest-frame EW of {\lya} $26\pm
 4\mbox{\AA}$) the outflows are ubiquitous at lower-$z$. At
 $z\sim 3$, \citet{erb14} reported Lyman-$\alpha$ velocity shifts of
 $\Delta{v}\approx100 - 500 \kms$ for low-EW (EW $\lesssim10\mbox{\AA}$)
 LAEs and a strong anti-correlation between $\Delta{v}$ and EW.  They
 concluded that $\Delta v$ is likely modulated both by galaxy continuum luminosity and by Lyman-$\alpha$ EW. At $z\sim 7$, however,
 \citet{stark15} measured an offset of
 $\Delta{v}\approx60~\rm{km}~s^{-1}$ between Lyman-$\alpha$ and CIV
 for a lensed low-luminosity galaxy A1703-zd6 at $z=7.045$
   (though CIV might not trace systemic velocity). \citet{pentericci16} reported velocity shifts of $\sim
 100\mbox{km/s}$ between {\ct} and {\lya} for their 3 most significant
 detections. For a higher luminosity galaxy  \citet{stark17} reported an offset between
 CIII] and {\lya} of $\sim 340\mbox{km/s}$. Given the small velocity offset of RXJ1347:1206
 (and similar other low luminosity galaxies) it seems that at $z\sim
 7$, {\lya} is much closer to systemic velocity than is the case for low-$z$ LAEs
 at similar UV-continuum luminosities.

 This is important, because  velocity offsets are crucial in
 interpreting the line visibility during the reionization epoch.  The
 low offset is interpreted differently in the so-called shell models
 vs. multi phase models of LAEs. A shell model requires low neutral
 gas column density or a high outflow velocity \citep{verhamme15}. In
 the case of a multi-phase model \citep{dijkstra16}, a low velocity
 offset translates to a low covering fraction of neutral gas,
 independent of its neutral column density and outflow
 velocity. In both models, however, the low velocity offset is a consequence of the presence of low HI-column density escape routes for {\lya} photons, which may also allow ionizing photons to escape (see \citealp{verhamme15,dijkstra16,verhamme16} for more detailed discussions).
 For a more general conclusion a larger
 sample is needed, but if future observations systematically 
 show smaller velocity offsets, this would imply that the
 observed drop in LAE fraction between $z\sim 6$ and 7
 (e.g. \citealp{schenker12,schenker14,tilvi14,caruana14,pentericci14,schmidt16}) is more easily
 explained by changes in the IGM, than in the circumgalactic medium or galactic intrinsic properties \citep{dijkstra14,mesinger15,choudhury15}.


\section{Conclusions} \label{sec:conclusions}

In this paper, we report a {\ct} ALMA detection of a low-luminosity
galaxy at $z=6.7655$. The {\ct} redshift agrees with {\lya} redshift
and the position agrees with the optical/UV {\hst} counterpart of this
object (within uncertainties). This indicates that the {\lya} is at
(or close to) resonant frequency, potentially requiring a lower IGM
neutral fraction to explain the drop in the LAE fraction from $z\sim7$
to $z\sim6$ if the kinematics of the {\lya} emitting region in
RXJ1347:1216 are typical of $z\sim7$ sub-$L^{\ast}$ galaxies.

The {\ct} luminosity is much lower than that expected from
low-redshift low-metallicity dwarf galaxies, yet it is consistent with
predictions of simulations of $Z<0.2Z_{\odot}$ galaxies at $z\sim
7$ by \citet{vallini15}. The departure of high-z galaxies from local relation is the most likely explanation for why
several searches of {\ct} emission at $z\sim 7$ have yielded non-detections, as many
have assumed the $L_{\rm{[C II]}}$-SFR relation for local dwarf galaxies \citep{delooze14}. Due to gravitational lensing we reach
lower intrinsic flux limits (factor of $\sim 5$) than similar
observations of field galaxies, and as a result we are able to study a
source that belongs to the bulk of the population at $z\sim 7$. As
shown in this study, such lens-magnified observations enable studies
of the ISM in the sources responsible for reionization. Future
high-resolution observations with ALMA will allow us to resolve
(spectrally and spatially) more galaxies at $z\sim7$, and study in
detail the kinematics and spatial distributions of {\ct}-emitting gas at sub-kpc scales. Such studies
will firmly establish ALMA as the premiere facility that will
revolutionize the explorations of the earliest galaxies and our
understanding of their place in the galaxy evolution and reionization
puzzle.

\acknowledgements We would like to thank Sergio Martin for help with
the ALMA data. The work was carried out by MB while visiting Joint
ALMA Observatory (JAO); MB acknowledges support for the visit by JAO
through ALMA visitor programe. This paper makes use of the following
ALMA data: ADS/JAO.ALMA\#2015.1.00091.S. ALMA is a partnership of ESO
(representing its member states), NSF (USA) and NINS (Japan), together
with NRC (Canada), NSC and ASIAA (Taiwan), and KASI (Republic of
Korea), in cooperation with the Republic of Chile. The Joint ALMA
Observatory is operated by ESO, AUI/NRAO and NAOJ. Support was
provided by NRAO via the NRAO Student Observing Support (SOS)
Program. This material is based upon work supported by Associated
Universities, Inc./National Radio Astronomy Observatory and the
National Science Foundation Observations were also carried out using
{\Spitzer} Space Telescope, which is operated by the Jet Propulsion
Laboratory, California Institute of Technology under a contract with
NASA. Also based on observations made with the NASA/ESA Hubble Space
Telescope, obtained at the Space Telescope Science Institute, which is
operated by the Association of Universities for Research in Astronomy,
Inc., under NASA contract NAS 5-26555 and NNX08AD79G and W.M. Keck
Observatory, which is operated as a scientific partnership among the
California Institute of Technology, the University of California and
the National Aeronautics and Space Administration. The Observatory was
made possible by the generous financial support of the W.M. Keck
Foundation.  Support for this work was provided by NASA through an
award issued by JPL/Caltech and through {\hst}-AR-13235,
{\hst}-GO-13459, {\hst}-GO-13177, {\hst}-GO-10200, {\hst}-GO-10863,
and {\hst}-GO-11099 from STScI. TT acknowledges support by the Packard
Fellowship.  AH acknowledges support by NASA Headquarters under the
NASA Earth and Space Science Fellowship Program Grant
ASTRO14F-0007. The authors wish to recognize and acknowledge the very
significant cultural role and reverence that the summit of Mauna Kea
has always had within the indigenous Hawaiian community.  We are most
fortunate to have the opportunity to conduct observations from this
mountain.

{\it Facilities:} \facility{ALMA}, \facility{Spitzer
  (IRAC)}, \facility{HST (ACS/WFC3)}, \facility{Keck:Deimos}

\bibliographystyle{apj}

\begin{thebibliography}{51}
\expandafter\ifx\csname natexlab\endcsname\relax\def\natexlab#1{#1}\fi

\bibitem[{{Bertin} \& {Arnouts}(1996)}]{sextractor}
{Bertin}, E. \& {Arnouts}, S. 1996, \aaps, 117, 393

\bibitem[{{Bouwens} {et~al.}(2015){Bouwens}, {Illingworth}, {Oesch}, {Trenti},
  {Labb{\'e}}, {Bradley}, {Carollo}, {van Dokkum}, {Gonzalez}, {Holwerda},
  {Franx}, {Spitler}, {Smit}, \& {Magee}}]{bouwens15}
{Bouwens}, R.~J., {Illingworth}, G.~D., {Oesch}, P.~A., {Trenti}, M.,
  {Labb{\'e}}, I., {Bradley}, L., {Carollo}, M., {van Dokkum}, P.~G.,
  {Gonzalez}, V., {Holwerda}, B., {Franx}, M., {Spitler}, L., {Smit}, R., \&
  {Magee}, D. 2015, \apj, 803, 34

\bibitem[{{Brada{\v c}} {et~al.}(2014){Brada{\v c}}, {Ryan}, {Casertano},
  {Huang}, {Lemaux}, {Schrabback}, {Gonzalez}, {Allen}, {Cain}, {Gladders},
  {Hall}, {Hildebrandt}, {Hinz}, {von der Linden}, {Lubin}, {Treu}, \&
  {Zaritsky}}]{surfsup}
{Brada{\v c}}, M., {Ryan}, R., {Casertano}, S., {Huang}, K.-H., {Lemaux},
  B.~C., {Schrabback}, T., {Gonzalez}, A.~H., {Allen}, S., {Cain}, B.,
  {Gladders}, M., {Hall}, N., {Hildebrandt}, H., {Hinz}, J., {von der Linden},
  A., {Lubin}, L., {Treu}, T., \& {Zaritsky}, D. 2014, \apj, 785, 108

\bibitem[{{Bradley} {et~al.}(2014){Bradley}, {Zitrin}, {Coe}, {Bouwens},
  {Postman}, {Balestra}, {Grillo}, {Monna}, {Rosati}, {Seitz}, {Host}, {Lemze},
  {Moustakas}, {Moustakas}, {Shu}, {Zheng}, {Broadhurst}, {Carrasco}, {Jouvel},
  {Koekemoer}, {Medezinski}, {Meneghetti}, {Nonino}, {Smit}, {Umetsu},
  {Bartelmann}, {Ben{\'{\i}}tez}, {Donahue}, {Ford}, {Infante}, {Jimenez-Teja},
  {Kelson}, {Lahav}, {Maoz}, {Melchior}, {Merten}, \& {Molino}}]{bradley14}
{Bradley}, L.~D., {Zitrin}, A., {Coe}, D., {Bouwens}, R., {Postman}, M.,
  {Balestra}, I., {Grillo}, C., {Monna}, A., {Rosati}, P., {Seitz}, S., {Host},
  O., {Lemze}, D., {Moustakas}, J., {Moustakas}, L.~A., {Shu}, X., {Zheng}, W.,
  {Broadhurst}, T., {Carrasco}, M., {Jouvel}, S., {Koekemoer}, A.,
  {Medezinski}, E., {Meneghetti}, M., {Nonino}, M., {Smit}, R., {Umetsu}, K.,
  {Bartelmann}, M., {Ben{\'{\i}}tez}, N., {Donahue}, M., {Ford}, H., {Infante},
  L., {Jimenez-Teja}, Y., {Kelson}, D., {Lahav}, O., {Maoz}, D., {Melchior},
  P., {Merten}, J., \& {Molino}, A. 2014, \apj, 792, 76

\bibitem[{{Bruzual} \& {Charlot}(2003)}]{bc03}
{Bruzual}, G. \& {Charlot}, S. 2003, \mnras, 344, 1000

\bibitem[{{Calzetti} {et~al.}(2000){Calzetti}, {Armus}, {Bohlin}, {Kinney},
  {Koornneef}, \& {Storchi-Bergmann}}]{calzetti00}
{Calzetti}, D., {Armus}, L., {Bohlin}, R.~C., {Kinney}, A.~L., {Koornneef}, J.,
  \& {Storchi-Bergmann}, T. 2000, \apj, 533, 682

\bibitem[{{Capak} {et~al.}(2015){Capak}, {Carilli}, {Jones}, {Casey},
  {Riechers}, {Sheth}, {Carollo}, {Ilbert}, {Karim}, {Lefevre}, {Lilly},
  {Scoville}, {Smolcic}, \& {Yan}}]{capak15}
{Capak}, P.~L., {Carilli}, C., {Jones}, G., {Casey}, C.~M., {Riechers}, D.,
  {Sheth}, K., {Carollo}, C.~M., {Ilbert}, O., {Karim}, A., {Lefevre}, O.,
  {Lilly}, S., {Scoville}, N., {Smolcic}, V., \& {Yan}, L. 2015, \nat, 522, 455

\bibitem[{{Carilli} \& {Walter}(2013)}]{carilli13}
{Carilli}, C. \& {Walter}, F. 2013, ArXiv:1301.0371

\bibitem[{{Caruana} {et~al.}(2014){Caruana}, {Bunker}, {Wilkins}, {Stanway},
  {Lorenzoni}, {Jarvis}, \& {Ebert}}]{caruana14}
{Caruana}, J., {Bunker}, A.~J., {Wilkins}, S.~M., {Stanway}, E.~R.,
  {Lorenzoni}, S., {Jarvis}, M.~J., \& {Ebert}, H. 2014, \mnras, 443, 2831

\bibitem[{{Choudhury} {et~al.}(2015){Choudhury}, {Puchwein}, {Haehnelt}, \&
  {Bolton}}]{choudhury15}
{Choudhury}, T.~R., {Puchwein}, E., {Haehnelt}, M.~G., \& {Bolton}, J.~S. 2015,
  \mnras, 452, 261

\bibitem[{{Cormier} {et~al.}(2015){Cormier}, {Madden}, {Lebouteiller}, {Abel},
  {Hony}, {Galliano}, {R{\'e}my-Ruyer}, {Bigiel}, {Baes}, {Boselli},
  {Chevance}, {Cooray}, {De Looze}, {Doublier}, {Galametz}, {Hughes},
  {Karczewski}, {Lee}, {Lu}, \& {Spinoglio}}]{cormier15}
{Cormier}, D., {Madden}, S.~C., {Lebouteiller}, V., {Abel}, N., {Hony}, S.,
  {Galliano}, F., {R{\'e}my-Ruyer}, A., {Bigiel}, F., {Baes}, M., {Boselli},
  A., {Chevance}, M., {Cooray}, A., {De Looze}, I., {Doublier}, V., {Galametz},
  M., {Hughes}, T., {Karczewski}, O.~{\L}., {Lee}, M.-Y., {Lu}, N., \&
  {Spinoglio}, L. 2015, \aap, 578, A53

\bibitem[{{De Looze} {et~al.}(2014){De Looze}, {Cormier}, {Lebouteiller},
  {Madden}, {Baes}, {Bendo}, {Boquien}, {Boselli}, {Clements}, {Cortese},
  {Cooray}, {Galametz}, {Galliano}, {Graci{\'a}-Carpio}, {Isaak}, {Karczewski},
  {Parkin}, {Pellegrini}, {R{\'e}my-Ruyer}, {Spinoglio}, {Smith}, \&
  {Sturm}}]{delooze14}
{De Looze}, I., {Cormier}, D., {Lebouteiller}, V., {Madden}, S., {Baes}, M.,
  {Bendo}, G.~J., {Boquien}, M., {Boselli}, A., {Clements}, D.~L., {Cortese},
  L., {Cooray}, A., {Galametz}, M., {Galliano}, F., {Graci{\'a}-Carpio}, J.,
  {Isaak}, K., {Karczewski}, O.~{\L}., {Parkin}, T.~J., {Pellegrini}, E.~W.,
  {R{\'e}my-Ruyer}, A., {Spinoglio}, L., {Smith}, M.~W.~L., \& {Sturm}, E.
  2014, \aap, 568, A62

\bibitem[{{Dijkstra}(2014)}]{dijkstra14}
{Dijkstra}, M. 2014, \pasa, 31, e040

\bibitem[{{Dijkstra} {et~al.}(2016){Dijkstra}, {Gronke}, \&
  {Venkatesan}}]{dijkstra16}
{Dijkstra}, M., {Gronke}, M., \& {Venkatesan}, A. 2016, \apj, 828, 71

\bibitem[{{Dunlop} {et~al.}(2016){Dunlop}, {McLure}, {Biggs}, {Geach},
  {Michalowski}, {Ivison}, {Rujopakarn}, {van Kampen}, {Kirkpatrick}, {Pope},
  {Scott}, {Swinbank}, {Targett}, {Aretxaga}, {Austermann}, {Best}, {Bruce},
  {Chapin}, {Charlot}, {Cirasuolo}, {Coppin}, {Ellis}, {Finkelstein},
  {Hayward}, {Hughes}, {Ibar}, {Khochfar}, {Koprowski}, {Narayanan},
  {Papovich}, {Peacock}, {Robertson}, {Vernstrom}, {van der Werf}, {Wilson}, \&
  {Yun}}]{dunlop16}
{Dunlop}, J.~S., {McLure}, R.~J., {Biggs}, A.~D., {Geach}, J.~E.,
  {Michalowski}, M.~J., {Ivison}, R.~J., {Rujopakarn}, W., {van Kampen}, E.,
  {Kirkpatrick}, A., {Pope}, A., {Scott}, D., {Swinbank}, A.~M., {Targett},
  T.~A., {Aretxaga}, I., {Austermann}, J.~E., {Best}, P.~N., {Bruce}, V.~A.,
  {Chapin}, E.~L., {Charlot}, S., {Cirasuolo}, M., {Coppin}, K.~E.~K., {Ellis},
  R.~S., {Finkelstein}, S.~L., {Hayward}, C.~C., {Hughes}, D.~H., {Ibar}, E.,
  {Khochfar}, S., {Koprowski}, M.~P., {Narayanan}, D., {Papovich}, C.,
  {Peacock}, J.~A., {Robertson}, B., {Vernstrom}, T., {van der Werf}, P.~P.,
  {Wilson}, G.~W., \& {Yun}, M. 2016, ArXiv:1606.00227

\bibitem[{{Erb} {et~al.}(2014){Erb}, {Steidel}, {Trainor}, {Bogosavljevi{\'c}},
  {Shapley}, {Nestor}, {Kulas}, {Law}, {Strom}, {Rudie}, {Reddy}, {Pettini},
  {Konidaris}, {Mace}, {Matthews}, \& {McLean}}]{erb14}
{Erb}, D.~K., {Steidel}, C.~C., {Trainor}, R.~F., {Bogosavljevi{\'c}}, M.,
  {Shapley}, A.~E., {Nestor}, D.~B., {Kulas}, K.~R., {Law}, D.~R., {Strom},
  A.~L., {Rudie}, G.~C., {Reddy}, N.~A., {Pettini}, M., {Konidaris}, N.~P.,
  {Mace}, G., {Matthews}, K., \& {McLean}, I.~S. 2014, \apj, 795, 33

\bibitem[{{Gonz{\'a}lez-L{\'o}pez} {et~al.}(2014){Gonz{\'a}lez-L{\'o}pez},
  {Riechers}, {Decarli}, {Walter}, {Vallini}, {Neri}, {Bertoldi}, {Bolatto},
  {Carilli}, {Cox}, {da Cunha}, {Ferrara}, {Gallerani}, \&
  {Infante}}]{gonzalezlopez14}
{Gonz{\'a}lez-L{\'o}pez}, J., {Riechers}, D.~A., {Decarli}, R., {Walter}, F.,
  {Vallini}, L., {Neri}, R., {Bertoldi}, F., {Bolatto}, A.~D., {Carilli},
  C.~L., {Cox}, P., {da Cunha}, E., {Ferrara}, A., {Gallerani}, S., \&
  {Infante}, L. 2014, \apj, 784, 99

\bibitem[{{Huang} {et~al.}(2016){Huang}, {Brada{\v c}}, {Lemaux}, {Ryan},
  {Hoag}, {Castellano}, {Amor{\'{\i}}n}, {Fontana}, {Brammer}, {Cain}, {Lubin},
  {Merlin}, {Schmidt}, {Schrabback}, {Treu}, {Gonzalez}, {von der Linden}, \&
  {Knight}}]{huang15a}
{Huang}, K.-H., {Brada{\v c}}, M., {Lemaux}, B.~C., {Ryan}, Jr., R.~E., {Hoag},
  A., {Castellano}, M., {Amor{\'{\i}}n}, R., {Fontana}, A., {Brammer}, G.~B.,
  {Cain}, B., {Lubin}, L.~M., {Merlin}, E., {Schmidt}, K.~B., {Schrabback}, T.,
  {Treu}, T., {Gonzalez}, A.~H., {von der Linden}, A., \& {Knight}, R.~I. 2016,
  \apj, 817, 11

\bibitem[{{Inoue} {et~al.}(2016){Inoue}, {Tamura}, {Matsuo}, {Mawatari},
  {Shimizu}, {Shibuya}, {Ota}, {Yoshida}, {Zackrisson}, {Kashikawa}, {Kohno},
  {Umehata}, {Hatsukade}, {Iye}, {Matsuda}, {Okamoto}, \&
  {Yamaguchi}}]{inoue16}
{Inoue}, A.~K., {Tamura}, Y., {Matsuo}, H., {Mawatari}, K., {Shimizu}, I.,
  {Shibuya}, T., {Ota}, K., {Yoshida}, N., {Zackrisson}, E., {Kashikawa}, N.,
  {Kohno}, K., {Umehata}, H., {Hatsukade}, B., {Iye}, M., {Matsuda}, Y.,
  {Okamoto}, T., \& {Yamaguchi}, Y. 2016, Science, 352, 1559

\bibitem[{{Kanekar} {et~al.}(2013){Kanekar}, {Wagg}, {Ram Chary}, \&
  {Carilli}}]{kanekar13}
{Kanekar}, N., {Wagg}, J., {Ram Chary}, R., \& {Carilli}, C.~L. 2013, \apjl,
  771, L20

\bibitem[{{Kennicutt}(1998)}]{kennicutt98}
{Kennicutt}, Jr., R.~C. 1998, \apj, 498, 541

\bibitem[{{Knudsen} {et~al.}(2016{\natexlab{a}}){Knudsen}, {Richard}, {Kneib},
  {Jauzac}, {Cl{\'e}ment}, {Drouart}, {Egami}, \& {Lindroos}}]{knudsen16}
{Knudsen}, K.~K., {Richard}, J., {Kneib}, J.-P., {Jauzac}, M., {Cl{\'e}ment},
  B., {Drouart}, G., {Egami}, E., \& {Lindroos}, L. 2016{\natexlab{a}}, \mnras,
  462, L6

\bibitem[{{Knudsen} {et~al.}(2016{\natexlab{b}}){Knudsen}, {Watson}, {Frayer},
  {Christensen}, {Gallazzi}, {Michalowski}, {Richard}, \&
  {Zavala}}]{knudsen16a}
{Knudsen}, K.~K., {Watson}, D., {Frayer}, D., {Christensen}, L., {Gallazzi},
  A., {Michalowski}, M.~J., {Richard}, J., \& {Zavala}, J. 2016{\natexlab{b}},
  ArXiv:1603.03222

\bibitem[{{Madau} \& {Haardt}(2015)}]{madau15}
{Madau}, P. \& {Haardt}, F. 2015, \apjl, 813, L8

\bibitem[{{Maiolino} {et~al.}(2015){Maiolino}, {Carniani}, {Fontana},
  {Vallini}, {Pentericci}, {Ferrara}, {Vanzella}, {Grazian}, {Gallerani},
  {Castellano}, {Cristiani}, {Brammer}, {Santini}, {Wagg}, \&
  {Williams}}]{maiolino15}
{Maiolino}, R., {Carniani}, S., {Fontana}, A., {Vallini}, L., {Pentericci}, L.,
  {Ferrara}, A., {Vanzella}, E., {Grazian}, A., {Gallerani}, S., {Castellano},
  M., {Cristiani}, S., {Brammer}, G., {Santini}, P., {Wagg}, J., \& {Williams},
  R. 2015, ArXiv:1502.06634

\bibitem[{{Mesinger} {et~al.}(2015){Mesinger}, {Aykutalp}, {Vanzella},
  {Pentericci}, {Ferrara}, \& {Dijkstra}}]{mesinger15}
{Mesinger}, A., {Aykutalp}, A., {Vanzella}, E., {Pentericci}, L., {Ferrara},
  A., \& {Dijkstra}, M. 2015, \mnras, 446, 566

\bibitem[{{Narayanan} \& {Krumholz}(2016)}]{narayanan16}
{Narayanan}, D. \& {Krumholz}, M. 2016, ArXiv:1601.05803

\bibitem[{{Olsen} {et~al.}(2015){Olsen}, {Greve}, {Narayanan}, {Thompson},
  {Toft}, \& {Brinch}}]{olsen15}
{Olsen}, K.~P., {Greve}, T.~R., {Narayanan}, D., {Thompson}, R., {Toft}, S., \&
  {Brinch}, C. 2015, \apj, 814, 76

\bibitem[{{Ota} {et~al.}(2014){Ota}, {Walter}, {Ohta}, {Hatsukade}, {Carilli},
  {da Cunha}, {Gonz{\'a}lez-L{\'o}pez}, {Decarli}, {Hodge}, {Nagai}, {Egami},
  {Jiang}, {Iye}, {Kashikawa}, {Riechers}, {Bertoldi}, {Cox}, {Neri}, \&
  {Weiss}}]{ota14}
{Ota}, K., {Walter}, F., {Ohta}, K., {Hatsukade}, B., {Carilli}, C.~L., {da
  Cunha}, E., {Gonz{\'a}lez-L{\'o}pez}, J., {Decarli}, R., {Hodge}, J.~A.,
  {Nagai}, H., {Egami}, E., {Jiang}, L., {Iye}, M., {Kashikawa}, N.,
  {Riechers}, D.~A., {Bertoldi}, F., {Cox}, P., {Neri}, R., \& {Weiss}, A.
  2014, \apj, 792, 34

\bibitem[{{Ouchi} {et~al.}(2013){Ouchi}, {Ellis}, {Ono}, {Nakanishi}, {Kohno},
  {Momose}, {Kurono}, {Ashby}, {Shimasaku}, {Willner}, {Fazio}, {Tamura}, \&
  {Iono}}]{ouchi13}
{Ouchi}, M., {Ellis}, R., {Ono}, Y., {Nakanishi}, K., {Kohno}, K., {Momose},
  R., {Kurono}, Y., {Ashby}, M.~L.~N., {Shimasaku}, K., {Willner}, S.~P.,
  {Fazio}, G.~G., {Tamura}, Y., \& {Iono}, D. 2013, \apj, 778, 102

\bibitem[{{Pallottini} {et~al.}(2016){Pallottini}, {Ferrara}, {Gallerani},
  {Vallini}, {Maiolino}, \& {Salvadori}}]{pallottini16}
{Pallottini}, A., {Ferrara}, A., {Gallerani}, S., {Vallini}, L., {Maiolino},
  R., \& {Salvadori}, S. 2016, ArXiv:1609.01719

\bibitem[{{Pei}(1992)}]{pei92}
{Pei}, Y.~C. 1992, \apj, 395, 130

\bibitem[{{Pentericci} {et~al.}(2016){Pentericci}, {Carniani}, {Castellano},
  {Fontana}, {Maiolino}, {Guaita}, {Vanzella}, {Grazian}, {Santini}, {Yan},
  {Cristiani}, {Conselice}, {Giavalisco}, {Hathi}, \&
  {Koekemoer}}]{pentericci16}
{Pentericci}, L., {Carniani}, S., {Castellano}, M., {Fontana}, A., {Maiolino},
  R., {Guaita}, L., {Vanzella}, E., {Grazian}, A., {Santini}, P., {Yan}, H.,
  {Cristiani}, S., {Conselice}, C., {Giavalisco}, M., {Hathi}, N., \&
  {Koekemoer}, A. 2016, ArXiv:1608.08837

\bibitem[{{Pentericci} {et~al.}(2014){Pentericci}, {Vanzella}, {Fontana},
  {Castellano}, {Treu}, {Mesinger}, {Dijkstra}, {Grazian}, {Brada{\v c}},
  {Conselice}, {Cristiani}, {Dunlop}, {Galametz}, {Giavalisco}, {Giallongo},
  {Koekemoer}, {McLure}, {Maiolino}, {Paris}, \& {Santini}}]{pentericci14}
{Pentericci}, L., {Vanzella}, E., {Fontana}, A., {Castellano}, M., {Treu}, T.,
  {Mesinger}, A., {Dijkstra}, M., {Grazian}, A., {Brada{\v c}}, M.,
  {Conselice}, C., {Cristiani}, S., {Dunlop}, J., {Galametz}, A., {Giavalisco},
  M., {Giallongo}, E., {Koekemoer}, A., {McLure}, R., {Maiolino}, R., {Paris},
  D., \& {Santini}, P. 2014, \apj, 793, 113

\bibitem[{{Postman} {et~al.}(2012){Postman}, {Coe}, {Ben{\'{\i}}tez},
  {Bradley}, {Broadhurst}, {Donahue}, {Ford}, {Graur}, {Graves}, {Jouvel},
  {Koekemoer}, {Lemze}, {Medezinski}, {Molino}, {Moustakas}, {Ogaz}, {Riess},
  {Rodney}, {Rosati}, {Umetsu}, {Zheng}, {Zitrin}, {Bartelmann}, {Bouwens},
  {Czakon}, {Golwala}, {Host}, {Infante}, {Jha}, {Jimenez-Teja}, {Kelson},
  {Lahav}, {Lazkoz}, {Maoz}, {McCully}, {Melchior}, {Meneghetti}, {Merten},
  {Moustakas}, {Nonino}, {Patel}, {Reg{\"o}s}, {Sayers}, {Seitz}, \& {Van der
  Wel}}]{postman12}
{Postman}, M., {Coe}, D., {Ben{\'{\i}}tez}, N., {Bradley}, L., {Broadhurst},
  T., {Donahue}, M., {Ford}, H., {Graur}, O., {Graves}, G., {Jouvel}, S.,
  {Koekemoer}, A., {Lemze}, D., {Medezinski}, E., {Molino}, A., {Moustakas},
  L., {Ogaz}, S., {Riess}, A., {Rodney}, S., {Rosati}, P., {Umetsu}, K.,
  {Zheng}, W., {Zitrin}, A., {Bartelmann}, M., {Bouwens}, R., {Czakon}, N.,
  {Golwala}, S., {Host}, O., {Infante}, L., {Jha}, S., {Jimenez-Teja}, Y.,
  {Kelson}, D., {Lahav}, O., {Lazkoz}, R., {Maoz}, D., {McCully}, C.,
  {Melchior}, P., {Meneghetti}, M., {Merten}, J., {Moustakas}, J., {Nonino},
  M., {Patel}, B., {Reg{\"o}s}, E., {Sayers}, J., {Seitz}, S., \& {Van der
  Wel}, A. 2012, \apjs, 199, 25

\bibitem[{{Robertson} {et~al.}(2015){Robertson}, {Ellis}, {Furlanetto}, \&
  {Dunlop}}]{robertson15}
{Robertson}, B.~E., {Ellis}, R.~S., {Furlanetto}, S.~R., \& {Dunlop}, J.~S.
  2015, \apjl, 802, L19

\bibitem[{{Schaerer} {et~al.}(2015){Schaerer}, {Boone}, {Zamojski}, {Staguhn},
  {Dessauges-Zavadsky}, {Finkelstein}, \& {Combes}}]{schaerer15}
{Schaerer}, D., {Boone}, F., {Zamojski}, M., {Staguhn}, J.,
  {Dessauges-Zavadsky}, M., {Finkelstein}, S., \& {Combes}, F. 2015, \aap, 574,
  A19

\bibitem[{{Schenker} {et~al.}(2014){Schenker}, {Ellis}, {Konidaris}, \&
  {Stark}}]{schenker14}
{Schenker}, M.~A., {Ellis}, R.~S., {Konidaris}, N.~P., \& {Stark}, D.~P. 2014,
  ArXiv:1404.4632

\bibitem[{{Schenker} {et~al.}(2012){Schenker}, {Stark}, {Ellis}, {Robertson},
  {Dunlop}, {McLure}, {Kneib}, \& {Richard}}]{schenker12}
{Schenker}, M.~A., {Stark}, D.~P., {Ellis}, R.~S., {Robertson}, B.~E.,
  {Dunlop}, J.~S., {McLure}, R.~J., {Kneib}, J.-P., \& {Richard}, J. 2012,
  \apj, 744, 179

\bibitem[{{Schmidt} {et~al.}(2016){Schmidt}, {Treu}, {Brada{\v c}}, {Vulcani},
  {Huang}, {Hoag}, {Maseda}, {Guaita}, {Pentericci}, {Brammer}, {Dijkstra},
  {Dressler}, {Fontana}, {Henry}, {Jones}, {Mason}, {Trenti}, \&
  {Wang}}]{schmidt16}
{Schmidt}, K.~B., {Treu}, T., {Brada{\v c}}, M., {Vulcani}, B., {Huang}, K.-H.,
  {Hoag}, A., {Maseda}, M., {Guaita}, L., {Pentericci}, L., {Brammer}, G.~B.,
  {Dijkstra}, M., {Dressler}, A., {Fontana}, A., {Henry}, A.~L., {Jones},
  T.~A., {Mason}, C., {Trenti}, M., \& {Wang}, X. 2016, \apj, 818, 38

\bibitem[{{Smit} {et~al.}(2014){Smit}, {Bouwens}, {Labb{\'e}}, {Zheng},
  {Bradley}, {Donahue}, {Lemze}, {Moustakas}, {Umetsu}, {Zitrin}, {Coe},
  {Postman}, {Gonzalez}, {Bartelmann}, {Ben{\'{\i}}tez}, {Broadhurst}, {Ford},
  {Grillo}, {Infante}, {Jimenez-Teja}, {Jouvel}, {Kelson}, {Lahav}, {Maoz},
  {Medezinski}, {Melchior}, {Meneghetti}, {Merten}, {Molino}, {Moustakas},
  {Nonino}, {Rosati}, \& {Seitz}}]{smit14}
{Smit}, R., {Bouwens}, R.~J., {Labb{\'e}}, I., {Zheng}, W., {Bradley}, L.,
  {Donahue}, M., {Lemze}, D., {Moustakas}, J., {Umetsu}, K., {Zitrin}, A.,
  {Coe}, D., {Postman}, M., {Gonzalez}, V., {Bartelmann}, M., {Ben{\'{\i}}tez},
  N., {Broadhurst}, T., {Ford}, H., {Grillo}, C., {Infante}, L.,
  {Jimenez-Teja}, Y., {Jouvel}, S., {Kelson}, D.~D., {Lahav}, O., {Maoz}, D.,
  {Medezinski}, E., {Melchior}, P., {Meneghetti}, M., {Merten}, J., {Molino},
  A., {Moustakas}, L.~A., {Nonino}, M., {Rosati}, P., \& {Seitz}, S. 2014,
  \apj, 784, 58

\bibitem[{{Stark} {et~al.}(2017){Stark}, {Ellis}, {Charlot}, {Chevallard},
  {Tang}, {Belli}, {Zitrin}, {Mainali}, {Gutkin}, {Vidal-Garc{\'{\i}}a},
  {Bouwens}, \& {Oesch}}]{stark17}
{Stark}, D.~P., {Ellis}, R.~S., {Charlot}, S., {Chevallard}, J., {Tang}, M.,
  {Belli}, S., {Zitrin}, A., {Mainali}, R., {Gutkin}, J.,
  {Vidal-Garc{\'{\i}}a}, A., {Bouwens}, R., \& {Oesch}, P. 2017, \mnras, 464,
  469

\bibitem[{{Stark} {et~al.}(2015){Stark}, {Walth}, {Charlot}, {Cl{\'e}ment},
  {Feltre}, {Gutkin}, {Richard}, {Mainali}, {Robertson}, {Siana}, {Tang}, \&
  {Schenker}}]{stark15}
{Stark}, D.~P., {Walth}, G., {Charlot}, S., {Cl{\'e}ment}, B., {Feltre}, A.,
  {Gutkin}, J., {Richard}, J., {Mainali}, R., {Robertson}, B., {Siana}, B.,
  {Tang}, M., \& {Schenker}, M. 2015, \mnras, 454, 1393

\bibitem[{{Tilvi} {et~al.}(2014){Tilvi}, {Papovich}, {Finkelstein}, {Long},
  {Song}, {Dickinson}, {Ferguson}, {Koekemoer}, {Giavalisco}, \&
  {Mobasher}}]{tilvi14}
{Tilvi}, V., {Papovich}, C., {Finkelstein}, S.~L., {Long}, J., {Song}, M.,
  {Dickinson}, M., {Ferguson}, H.~C., {Koekemoer}, A.~M., {Giavalisco}, M., \&
  {Mobasher}, B. 2014, \apj, 794, 5

\bibitem[{{Treu} {et~al.}(2015){Treu}, {Schmidt}, {Brammer}, {Vulcani}, {Wang},
  {Brada{\v c}}, {Dijkstra}, {Dressler}, {Fontana}, {Gavazzi}, {Henry}, {Hoag},
  {Huang}, {Jones}, {Kelly}, {Malkan}, {Mason}, {Pentericci}, {Poggianti},
  {Stiavelli}, {Trenti}, \& {von der Linden}}]{treu15}
{Treu}, T., {Schmidt}, K.~B., {Brammer}, G.~B., {Vulcani}, B., {Wang}, X.,
  {Brada{\v c}}, M., {Dijkstra}, M., {Dressler}, A., {Fontana}, A., {Gavazzi},
  R., {Henry}, A.~L., {Hoag}, A., {Huang}, K.-H., {Jones}, T.~A., {Kelly},
  P.~L., {Malkan}, M.~A., {Mason}, C., {Pentericci}, L., {Poggianti}, B.,
  {Stiavelli}, M., {Trenti}, M., \& {von der Linden}, A. 2015, \apj, 812, 114

\bibitem[{{Vallini} {et~al.}(2016){Vallini}, {Ferrara}, {Pallottini}, \&
  {Gallerani}}]{vallini16}
{Vallini}, L., {Ferrara}, A., {Pallottini}, A., \& {Gallerani}, S. 2016,
  ArXiv:1606.08464

\bibitem[{{Vallini} {et~al.}(2015){Vallini}, {Gallerani}, {Ferrara},
  {Pallottini}, \& {Yue}}]{vallini15}
{Vallini}, L., {Gallerani}, S., {Ferrara}, A., {Pallottini}, A., \& {Yue}, B.
  2015, \apj, 813, 36

\bibitem[{{Verhamme} {et~al.}(2015){Verhamme}, {Orlitov{\'a}}, {Schaerer}, \&
  {Hayes}}]{verhamme15}
{Verhamme}, A., {Orlitov{\'a}}, I., {Schaerer}, D., \& {Hayes}, M. 2015, \aap,
  578, A7

\bibitem[{{Verhamme} {et~al.}(2016){Verhamme}, {Orlitova}, {Schaerer},
  {Izotov}, {Worseck}, {Thuan}, \& {Guseva}}]{verhamme16}
{Verhamme}, A., {Orlitova}, I., {Schaerer}, D., {Izotov}, Y., {Worseck}, G.,
  {Thuan}, T.~X., \& {Guseva}, N. 2016, ArXiv:1609.03477

\bibitem[{{Wang} {et~al.}(2013){Wang}, {Wagg}, {Carilli}, {Walter}, {Lentati},
  {Fan}, {Riechers}, {Bertoldi}, {Narayanan}, {Strauss}, {Cox}, {Omont},
  {Menten}, {Knudsen}, {Neri}, \& {Jiang}}]{wang13}
{Wang}, R., {Wagg}, J., {Carilli}, C.~L., {Walter}, F., {Lentati}, L., {Fan},
  X., {Riechers}, D.~A., {Bertoldi}, F., {Narayanan}, D., {Strauss}, M.~A.,
  {Cox}, P., {Omont}, A., {Menten}, K.~M., {Knudsen}, K.~K., {Neri}, R., \&
  {Jiang}, L. 2013, \apj, 773, 44

\bibitem[{{Willott} {et~al.}(2015){Willott}, {Carilli}, {Wagg}, \&
  {Wang}}]{willot15}
{Willott}, C.~J., {Carilli}, C.~L., {Wagg}, J., \& {Wang}, R. 2015, \apj, 807,
  180

\end{thebibliography}

\end{document}